# Time-Resolved Measurements of Electron Density in Nanosecond Pulsed Plasmas Using Microwave Scattering

Xingxing Wang, Paul Stockett, Ravichandra Jagannath, Sally Bane, and Alexey Shashurin

*Purdue University, 701 W Stadium Avenue, West Lafayette, IN 47907*

## Abstract

In this work, Rayleigh microwave scattering was utilized to measure the electron number density produced by nanosecond high voltage breakdown in air between two electrodes in a pin-to-pin configuration (peak voltage 26 kV and pulse duration 55 ns). The peak electron density decreased from $1 \cdot 10^{17}$ cm$^{-3}$ down to $7 \cdot 10^{14}$ cm$^{-3}$ when increasing the gap distance from 2 to 8 mm (total electron number decreased from $2 \cdot 10^{13}$ down to $5 \cdot 10^{11}$ respectively). Electron number density decayed on the timescale of about several µs due to dissociative recombination.

Atmospheric pressure plasmas have unique characteristics which can be leveraged for a wide range of applications, including plasma-assisted ignition and combustion, aerodynamic flow control, medicine and environment, nanotechnology, and materials processing [1]. Detailed diagnostics of atmospheric pressure plasmas generated under various conditions (electrode configuration, discharge-driving high voltage waveform, background gas composition and temperature, etc.) remain a critical challenge. In recent years, nanosecond repetitively pulsed (NRP) plasma discharges have attracted great interest as an efficient method for creation of non-equilibrium plasmas at high pressures. The very large voltage rise over very short time scales and subsequent strong electric field accelerates electrons to very high energies, leading to efficient production of reactive and ionized species. Pulsing of these discharges at high repetition rates results in accumulation of metastable and active species. These plasmas are especially of interest in combustion and aerodynamic flow control applications. In the combustion field, NRP plasmas have been used to increase flame speed [2] [3], enhance flame stabilization at low fuel/oxidizer ratios [4] [5] [6] [7], and mitigate combustion instability [8] [9] [10] [11]. For aerodynamic flow control, surface dielectric barrier discharge (DBD) plasmas driven by NRP discharges has been explored for controlling boundary layers at realistic flight conditions [12] [13], and for shock wave modification in supersonic flow [14] [15]. NRP spark plasmas have also been shown to be effective at inducing flow instabilities by generating pressure waves [16] [17] [18].

High-resolution measurements of the temporal evolution of NRP plasmas are critical for understanding the discharge physics and ultimately for designing plasma sources tailored to specific applications. Conventional diagnostics of NRP discharges include optical emission spectroscopy (OES), Stark broadening, fast photography, measurements of voltage and current waveforms, etc. [19] [20] [21] [22] [23]. One of the most important characteristics of NRP plasmas is electron number density. However, direct experimental data available on electron number density and its evolution in time is limited and the measurement techniques used often have some significant limitations.

Electron density has been measured by several groups utilizing OES. Rusterholtz et al. estimated the peak electron density of NRP spark plasma in preheated (1000 K) air to be in the range of $10^{15}$ cm$^{-3}$ using spectra of the Balmer $H_\alpha$ line [19]. Van der Horst and co-workers, performed similar measurements in room temperature air and found the peak electron density to be approximately $3 \cdot 10^{18}$ cm$^{-3}$ [20]. In these studies, this method is restricted to the time range from 3-11 ns due to masking of the $H_\alpha$ line by $N_2$(B-A) emission after 11 ns and requires the background $N_2$(B-A) spectrum to be accurately determined and subtracted. Another method for measuring electron density from OES spectra is to use Stark broadening of the atomic nitrogen line and $H_\alpha/H_\beta$. This method was used to measure electron density of NRP plasmas in various kinds of gas mixtures, with resulting electron density was in the range of $10^{15}$ - $10^{18}$ cm$^{-3}$ [22] [23]. One potential issue, however, is that Stark broadening sensitivity is limited for electron densities > $10^{15}$ cm$^{-3}$ [23]. Furthermore, the interpretation of the broadening of the spectrum can be problematic due to contributions from several competing mechanisms to the line broadening.

In several detailed studies on NRP plasmas in preheated (1000 K) air with a pin-to-pin electrode configuration, Pai et al. identified three different plasma discharge regimes: corona, glow, and spark [24] [25] [26]. These regimes are categorized by the energy deposited per pulse: <10 µJ for corona, 10-100 µJ for diffuse or glow, and >100 µJ for spark. The resulting gas heating is <200 K for the corona and glow regimes [24] and in the range of 2000-4000 K for the spark regime [25]. The electron density in the spark mode was estimated to be in the range of $3\cdot10^{13}$ to $3\cdot10^{15}$ cm$^{-3}$ [26]. These values of electron density were estimated using measurements of the discharge voltage ($V_d$) and current ($I_d$). First, the voltage and current were used to calculate the equivalent resistance of the plasma column as $R = \frac{V_d}{I_d}$. Then, the plasma conductivity was determined based on the visually observed length ($L$) and cross-sectional area ($S$) of the plasma column as $\sigma = \frac{L}{R \cdot S}$. The plasma density was then calculated as $n_e = \frac{\sigma \cdot v \cdot m}{e^2}$ where $v$ is the electron-gas collision frequency, and $m$ and $e$ are the electron mass and charge, respectively. However, this approach to estimating electron density has some significant limitations. First, the voltage drop in the plasma column is different from $V_d$ (especially in the case of low-$V_d$ arc-type discharges) due to the presence of near electrode sheaths [27]. This causes overestimation of plasma column resistance and thus, underestimation of conductivity and plasma density. Second, the experimental data presented by the authors is not consistent with their assumption that the plasma acts as a purely resistive load. During the discharge, the times of zero-crossings of the $V_d$ and $I_d$ waveforms are shifted with respect to each other (e.g. see the figure 6 in Ref. [26] at $t \approx 10$ ns where $V_d \approx 0$ while $I_d \approx 30$ A). This offset in time between the discharge voltage and current clearly indicates the presence of a reactive part in the plasma load impedance and causes unreasonable estimation of the plasma density near the moment where $V_d \approx 0$ if method proposed by the authors' is used since $n_e \propto \sigma \propto \frac{1}{R} \propto \frac{I_d}{V_d} \to \infty$.

Thus, reliable diagnostics to measure the electron number density in the NRP plasmas are not available even though this parameter has critical importance for practical applications and validation of numerical codes. In this paper, a method for direct measurement of electron number density in the plasma column of the NRP discharge in a pin-to-pin electrode configuration by means of Rayleigh microwave scattering (RMS) is presented. This method was recently proposed and successfully applied for measurements in various types of atmospheric pressure microplasmas [28] [29] [30] [31] [32].

The plasma was generated between two pointed-tip tungsten electrodes separated by an inter-electrode gap distance *L*. In this work, three different gap distances were used: *L* = 2, 5, and 8 mm. Capacitance of the electrode geometry was measured to be around C=3.5 pF (by measuring electrode voltage and current waveform without plasma generation). The plasma was produced using high voltage pulses with pulse duration (FWHM) of 55 ns generated by a nanosecond high voltage pulser (Eagle Harbor NSP-3300-20-F). Short series of pulses (up to 10 pulses 1 ms apart) was used to ensure the plasma creation. The peak voltage applied to the pin-to-pin electrodes was approximately 26 kV. The voltage between the electrodes was measured by two high voltage probes (Tektronix P6015A) connected to the corresponding electrodes with grounds of both probes connected together to obtain a differential measurement to maintain the pulser output floating. Thus, the actual voltage applied to the electrodes was calculated as the difference between the measurements of the two voltage probes. The waveform of the voltage pulse sent to the electrodes with no breakdown of the gap is shown in Fig 1. Discharge current was measured by a high-bandwidth current transformer (Bergoz FCT-028-0.5-WB) on the positive voltage line. Time resolved images were taken by an ICCD camera (Princeton Instrument PI-MAX 1024i).

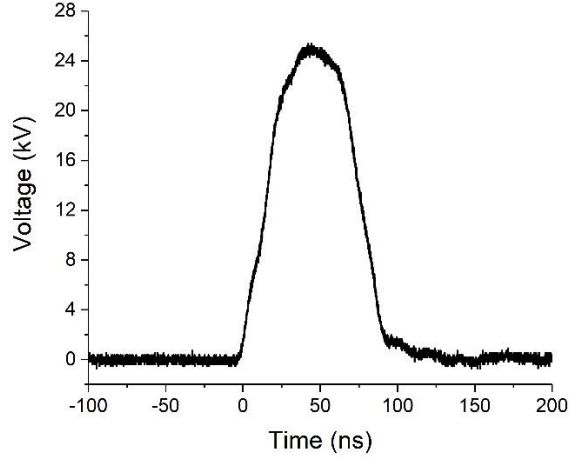

**Fig 1. Voltage pulse applied to the electrodes without breakdown of the electrode gap.**

A Rayleigh Microwave Scattering (RMS) system was constructed for measurement of the time-resolved plasma electron density. The RMS system utilizes a homodyne detection scheme for measurement of microwave signal scattered from the plasma volume. The system can be used to determine the total number of electrons inside of the plasma volume by analyzing the scattering of linearly polarized microwave radiation from the collinearly-oriented plasma channel. A schematic of the RMS system and its location in the overall experimental setup are shown in Fig 2. The microwave signal is radiated onto the testing object through a 10 dB horn antenna. The detected signal is received by another horn antenna and processed by an I/Q mixer outputting two signals: *I* and *Q*. The final output signal can be expressed as $U_{out} = \sqrt{I^2 + Q^2}$. The frequency of the microwave was set to be 10.9 GHz which minimized the noise level produced by the circuitry and surroundings.

Absolute calibration of the RMS system is accomplished using dielectric scatterers with known physical properties. A Teflon bullet ($\varepsilon = 2.1$) in the shape of a cylinder with a diameter of 0.32 cm and length of 1 cm was used in this work. The output $U_{out}$ of the RMS system can be related to properties of a dielectric scatterer and plasma as follows [28]:

$$U_{out} = \begin{cases} A\frac{e^2}{m\nu}N_e, & for\ plasma \\ AV\varepsilon_0(\varepsilon - 1)\omega, & for\ dielectric\ scatterers \end{cases} \quad (1)$$

where *A* is the proportionality coefficient, *e* is the electron charge, *v* is the electron-gas collision frequency which was estimated to be $1.46 \cdot 10^{12}$ sec$^{-1}$ in this case [28] [33], *V* is the volume of the dielectric scatterer, $\varepsilon_0$ is the vacuum permittivity, $\varepsilon$ is the relative permittivity of the dielectric material and $\omega$ is the angular frequency of the microwaves. Using the output measured for the Teflon bullet in the bottom relation in Eq. 1 the proportionality factor *A* was found to be 85.3 kVΩ/m². As a result, the top relation of Eqn 1 can be re-written as:

$$N_e = \frac{U_{out}[V]}{1.643 \cdot 10^{-15}} \quad (2)$$

and the total number of electrons can be calculated directly using the RMS output signal. The distance between both the radiating and receiving horn antennae and the testing object was kept at 6 cm to ensure flatness of the microwave front when hitting the plasma and to maintain the scattered signal within the Rayleigh scattering regime. More details on the RMS system can be found in [32].

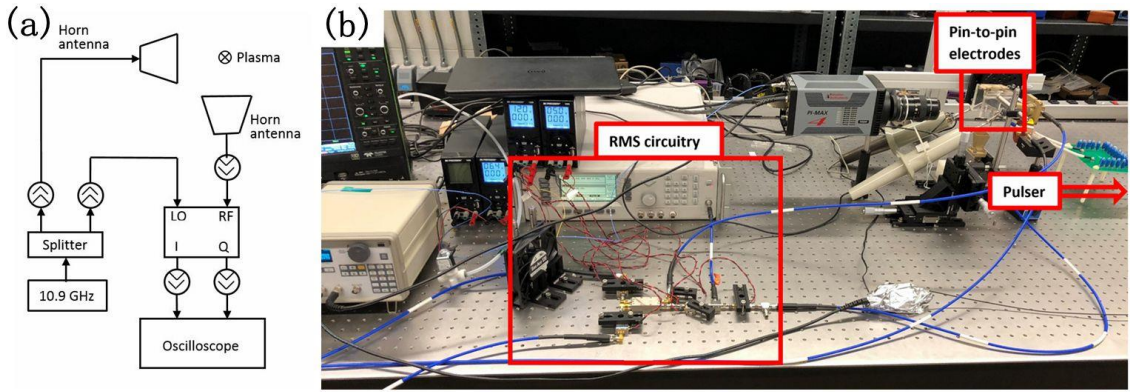

**Fig 2. (a) Schematic of the RMS system. (b) Entire experimental setup.**

Fig 3 shows the images of the NRP discharge for three different gap distances at different times after the breakdown (each image was taken at a different breakdown event). Analysis of the intensity distribution of ICCD images of the plasma column indicated that the plasma diameter was about 330 μm for all experimental conditions. Then, the volume of the plasma column was determined by assuming a cylindrical shape with height equal to gap distance *L* and diameter of 330 μm as $V_0[cm^3] = 8.6 \cdot 10^{-5} \cdot L[mm]$ and thus spatially averaged electron number density $n_e$ can be calculated as: $n_e = \frac{N_e}{V_0}$

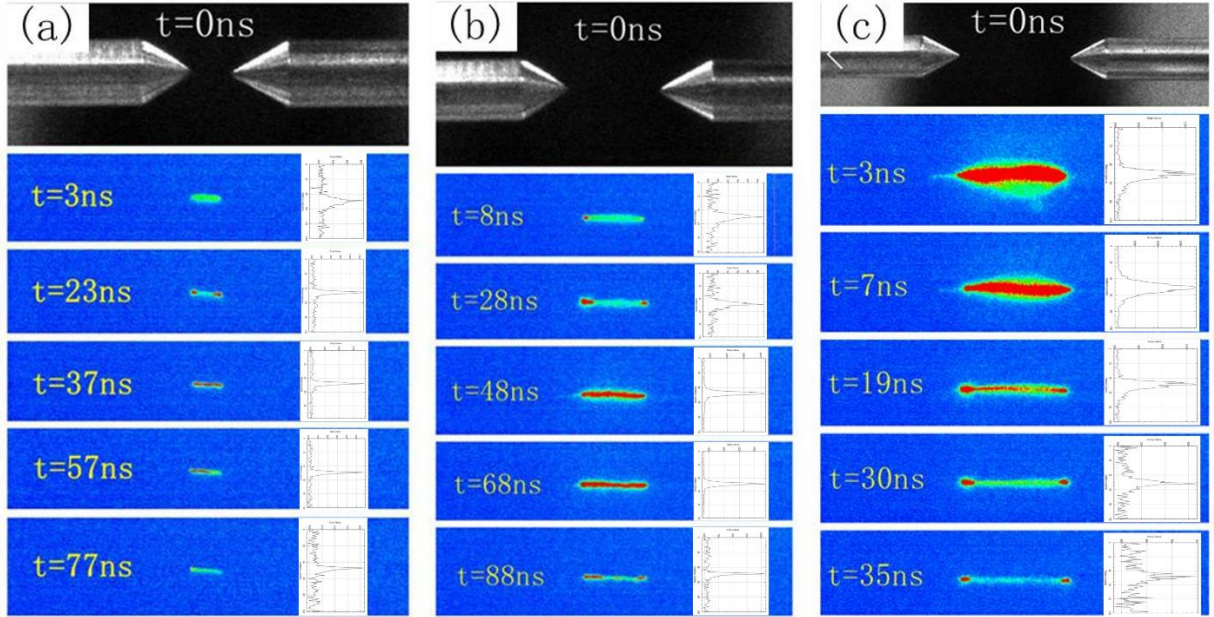

**Fig 3. Images of an NRP discharge plasma in a pin-to-pin electrode configuration with varying gap distances *L* at different moments of time after breakdown. The distribution of intensities for each case is plotted on the right-hand side of each ICCD image. (a) *L* = 2mm; (b) *L* = 5mm; (c) *L* = 8 mm. Exposure time = 3 ns.**

The temporal evolution of the average electron density $n_e$ determined from RMS measurements and the discharge voltage and current $L$ = 5 mm are plotted in Fig 4. Fig 4(a) shows the temporal evolution of $n_e$ on the microsecond timescale. The density reached its peak value of about $7 \cdot 10^{15}$ cm$^{-3}$ at the time of breakdown and then decayed in approximately 2 μs after the discharge initiation. Fig 4(b) shows the evolution of $N_e$ on a sub-microsecond timescale plotted along with the discharge voltage and current waveforms. Prior to electrical breakdown between the electrodes, the applied voltage increased until it reached the threshold value for breakdown at about 15 kV. Then the voltage rapidly decreased and current increased due to the breakdown and formation of a highly conductive plasma channel between the electrodes. Note, $I_d$ measurements show the discharge current only (obtained by subtraction of the displacement current from the total current measured by the current probe). The current reached a peak value of 15 A immediately after the breakdown and dropped down to zero at the end of the driving voltage pulse. Note that measurements of $V_d$ are presented for times before the breakdown only since limited bandwidth of the voltage probe (75 MHz) does not allow it to realistically capture the rapid drop of the voltage near the breakdown event. One can also see that $n_e$ decayed at a much slower rate compared to $I_d$. This indicates that plasma lifetime extends significantly longer than the duration of the discharge-driving pulse. This relatively long plasma decay is governed by dissociative recombination [26].

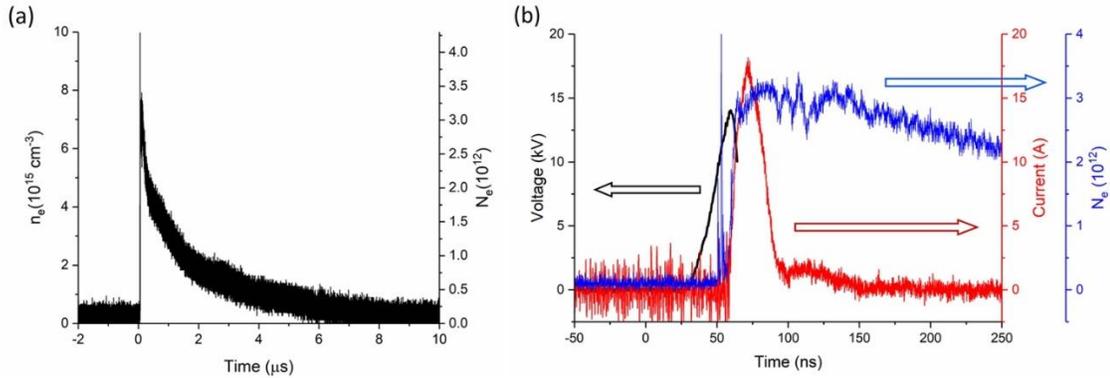

**Fig 4. (a)** Temporal evolution of $n_e$ and $N_e$ on the µs time scale. **(b)** Temporal evolution of voltage, current, and $N_e$ on the ns time scale.

Fig 5 shows the relationship between the peak value of electron density and total number of electrons in plasma volume versus gap distance $L$. As $L$ increases, $n_e$ and $N_e$ decreases, which can be possibly explained by a transition between discharge regimes similar to that discussed in Ref. [24].

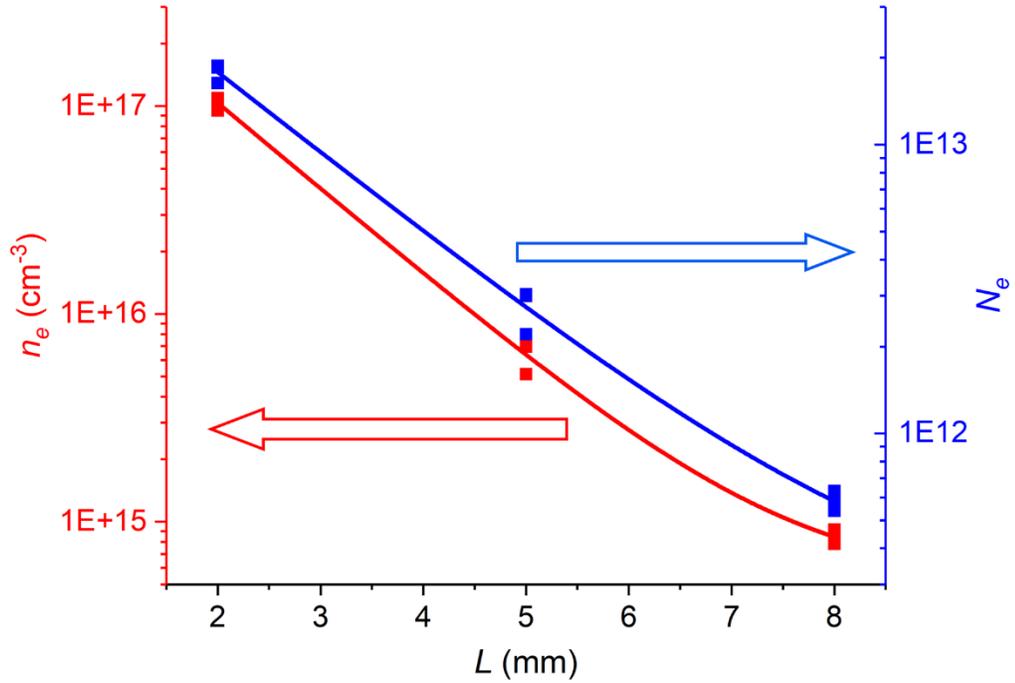

**Fig 5. peak value of $n_e$ and $N_e$ vs gap distance $L$**

To conclude, in this work, time-resolved measurement of electron number density in the plasma produced by nanosecond repetitive HV pulses using Rayleigh Microwave Scattering was demonstrated. The peak electron number density increased from $8·10^{14}$ to $1·10^{17}$ cm$^{-3}$ with value with decrease of inter-electrode gap distance from 8 to 2 mm ($N_e$ increased from $5·10^{11}$ to $2·10^{13}$, respectively). Efforts to determine the effect of pulse parameters and electrode geometry on the plasma regime and characteristics using RMS measurements are currently underway.

## Acknowledgements

We thank Dr. A. Garner, Dr. C. Scalo and Dr. P. Vlachos for useful discussions. This work was supported by U.S. Department of Energy (Grant No. DE-SC0018156).